\documentclass[%
 aip,
 amsmath,amssymb,
 reprint,%
]{revtex4-1}

\usepackage{graphicx}
\usepackage{dcolumn}
\usepackage{bm}

\usepackage[utf8]{inputenc}
\usepackage[T1]{fontenc}
\usepackage{mathptmx}
\usepackage{etoolbox}
\usepackage{hyperref}
\usepackage{physics}
\usepackage{rotating}
\usepackage{multirow}
\usepackage{threeparttable}

\newcommand\PM{\phantom{$-$}}

\makeatletter
\def\@email#1#2{%
 \endgroup
 \patchcmd{\titleblock@produce}
  {\frontmatter@RRAPformat}
  {\frontmatter@RRAPformat{\produce@RRAP{*#1\href{mailto:#2}{#2}}}\frontmatter@RRAPformat}
  {}{}
}%
\makeatother
\begin{document}

\preprint{AIP/123-QED}

\title[]{Benchmarking ionization potentials from the simple
pCCD model}
\author{Saddem Mamache}
 \altaffiliation[]{Institute of Physics, Faculty of Physics, Astronomy and Informatics, Nicolaus Copernicus University in Toru\'n, Grudziadzka 5, 87-100 Toru\'n, Poland.}
\author{Marta Ga\l{}y\'nska}%
\affiliation{ 
Institute of Physics, Faculty of Physics, Astronomy and Informatics, Nicolaus Copernicus University in Toru\'n, Grudziadzka 5, 87-100 Toru\'n, Poland.
}%
\author{Katharina Boguslawski}%
 \email{k.boguslawski@fizyka.umk.pl}
\affiliation{ 
Institute of Physics, Faculty of Physics, Astronomy and Informatics, Nicolaus Copernicus University in Toru\'n, Grudziadzka 5, 87-100 Toru\'n, Poland.
}

\date{\today}

\begin{abstract}
The electron-detachment energy is measured by its ionization potential (IP). As a result, it is a fundamental observable and important molecular electronic signature in photoelectron spectroscopy. A precise theoretical prediction of electron-detachment energies or ionization potentials is essential for organic optoelectronic systems like transistors, solar cells, or light-emitting diodes. In this work, we benchmark the performance of the recently presented IP variant of the equation-of-motion pair coupled cluster doubles (IP-EOM-pCCD) model to determine IPs. Specifically, the predicted ionization energies are compared to experimental results and higher-order coupled cluster theories based on statistically assessing 201 electron-detached states of 41 organic molecules for three different molecular orbital basis sets and two sets of particle-hole operators. While IP-EOM-pCCD features a reasonable spread and skewness of ionization energies, its mean error and standard deviation deviate up to 1.5 eV from reference data. Our study, thus, highlights the importance of dynamical correlation to reliably predict IPs from a pCCD reference function in small organic molecules.
\end{abstract}

\maketitle

\section{Introduction}
Quantum chemistry has devised a multitude of methods to study ionization and electron-detachment processes.
Examples include coupled cluster (CC) approaches,~\cite{paldus1992coupled,bartlett2007coupled,monika_mrcc}
density functional theory and its time-dependent formulation,~\cite{yang_parr_book}
configuration interaction models,~\cite{foster1960canonical,szalay2012}
perturbation theory,~\cite{shavitt_book}
algebraic-diagrammatic construction (ADC) schemes,~\cite{Schirmer1998adc,Dempwolff2019adc,Banerjee2019adc,Banerjee2021adc}
and Monte Carlo methods.~\cite{qmc,qmc-book-chapter}
The choice of method depends on the size and complexity of the system being studied, the level of accuracy required, and the availability of computational resources.
Well-established correlation methods for photoelectron spectroscopy simulations\cite{geertsen1989equation,stanton1993equation,watts1994inclusion,van2000benchmark,musial2003equation,krylov2008equation,manohar2008noniterative,cooper2010benchmark,evangelista2011alternative,musial2014equation,lischka2018multireference,gulania2021equation,marie2021variational}
are, for instance, ionization potential equation of motion coupled cluster (IP-EOM-CC) approaches.~\cite{nooijen1992coupled,nooijen1993coupled,Stanton1994-ip,Stanton1999-ip}
Among these, the most common flavors restrict the CC ansatz to single and double excitations (CCSD) or singles, doubles, and perturbative triples, where the latter results in the so-called gold standard of quantum chemistry CCSD(T).
In order to improve the performance of CCSD or CCSD(T), full triple, quadruple (Q), or higher-order excitations can be included in the CC ansatz, yielding the CCSDT, CCSDTQ, etc.~models, which are then combined with an IP-EOM formalism to describe ionized states.~\cite{musial2003equation}
In contrast to these well-established EOM-CC-based methods,~\cite{krylov2008equation,ee-cc-methods2012,eom-cc-bartlett2012} some of us recently presented the simplified IP-EOM pair coupled cluster doubles (IP-EOM-pCCD) variant~\cite{boguslawski2021open} as an inexpensive alternative to model open-shell electronic structures within the pCCD model.
A first numerical study~\cite{boguslawski2021open} demonstrated that the accuracy of IP-EOM-pCCD approaches the CCSD(T) or density matrix renormalization group~\cite{white,ors_springer,marti2010b,chanreview} (DMRG) level of accuracy in open-shell electronic structures.

pCCD~\cite{boguslawski2014efficient,stein2014seniority} was originally introduced as a geminal-based wavefunction ansatz.~\cite{limacher2013new}
Such wavefunction ansätze~\cite{hurley1953molecular,parr1956generalized,bardeen1957theory,parks1958theory,coleman1965structure,miller1968electron,tecmer2022geminal} provide a promising alternative to conventional state-of-the-art electronic structure approaches, which are typically constructed from one-electron functions to model nondynamic and static electron correlation effects.
In the realm of geminal-based methods, the strictly localized geminal (SLG) ansatz is widely used.~\cite{surjan1984interaction,poirier1987application,surjan1994interaction}
The antisymmetrized product of strongly orthogonal geminals~\cite{hurley1953molecular,parks1958theory,kutzelnigg1964direct,surjan1999introduction} and the generalized valence bond perfect pairing~\cite{bobrowicz1977self,cullen1996generalized} models are two well-known special cases of SLG, where the geminals are restricted to be strongly orthogonal,~\cite{arai1960theorem} which rules out intergeminal correlation.
In contrast, the pCCD ansatz is a geminal model that is computationally feasible while simultaneously taking into account intergeminal correlations.
It can be understood as a simple reduction of the single-reference CCD approach, where the cluster operator is constrained to electron-pair excitations only $\hat{T}_\textrm{pCCD}$, 
\begin{equation}
\ket{\textrm{pCCD}} = e^{\hat{T}_\textrm{pCCD}} \ket{\Phi_0},
\end{equation}
where
\begin{equation}
\hat{T}_\textrm{pCCD} = \sum_{i}^{n_{\rm occ}} \sum_{a}^{n_{\rm virt}} c_i^a{a_a^\dagger}{a_{\bar a}^\dagger}{a_{\bar i}}{a_i}.
\end{equation}
In the above equations, $\big|\Phi_0\rangle$ is some reference determinant, $\hat{a}_p$ ($\hat{a}_p^\dag$) are the elementary annihilation (creation) operators for spin-up $p$ and spin-down $(\overline{p})$ electrons, and $c^a_i$ are the pCCD cluster amplitudes, where the sum runs over all occupied $i$ and virtual $a$ orbitals.
Size extensivity is guaranteed by the exponential form of the ansatz, while size consistency is re-established through molecular orbital optimization.~\cite{boguslawski2014nonvariational,boguslawski2014projected,boguslawski2014efficient,stein2014seniority}
The resulting molecular orbital basis is localized and symmetry-broken and can be used to simulate quantum states with (quasi-)degeneracies.~\cite{boguslawski2016analysis}
Although being a simplification, the pCCD model proved to be a low-cost wavefunction ansatz for describing strongly-correlated closed-shell molecules.
Examples of frequent testing grounds for geminal-based models include the process of bond breaking and formation in small molecules~\cite{poirier1987application,rassolov2002geminal,small2012fusion,tecmer2014assessing,limacher2015orbital,tecmer2015singlet,tokmachev2016perspectives,brzek2019benchmarking,henderson2019geminal,nowak2021orbital,leszczyk2021assessing,leszczyk2022}
or heavy-element-containing compounds featuring actinide~\cite{tecmer2015singlet,garza2015actinide,boguslawski2016targeting,boguslawski2017erratum,nowak2019assessing,leszczyk2022,Nowak2023} or lanthanide~\cite{tecmer2019modeling} atoms.
Despite the growing interest in creating efficient and reliable geminal-based methods to account for various forms of electron correlation effects,~\cite{kallay1999improving,rosta2000interaction,rassolov2004geminal,zoboki2013linearized,pastorczak2015erpa,henderson2014seniority,limacher2014simple,Garza2015synergy,garza2015actinide,boguslawski2015linearized,chatterjee2016minimalistic,boguslawski2017benchmark,henderson2019geminal,Nowak2023} orbital- or density-based approaches are still widely employed in computational quantum chemistry.
Examples include the complete active space self-consistent-field approach\cite{Roos1987,Olsen1988} and the DMRG algorithm.

Although originally introduced as a closed-shell theory, the pCCD model has recently been extended to target open-shell molecular structures.~\cite{boguslawski2021open}
Open-shell electronic structures are created exploiting the IP-EOM formalism by removing electrons from the closed-shell pCCD reference function through a linear ansatz to parametrize the k-th state,\cite{boguslawski2016targeting,boguslawski2017erratum,boguslawski2018targeting,nowak2019assessing,boguslawski2021open}
\begin{equation}
\ket{\Psi_k} = \hat R(k)\ket{\textrm{pCCD}}
\end{equation}
where the operator $\hat R(k)$ generates the targeted state $k$ from the initial pCCD reference state.
The ionization operator $\hat R$ is typically divided into different parts based on the number of particle (electron creation) and hole (electron annihilation) operators contained in each component.
the single IP-EOM formalism defines $\hat R(k)$ as
\begin{equation}
\hat R^{\rm IP} = \sum_{i}r_i\hat a_i+ \frac{1}{2}\sum_{ij{a}}{r}^a_{ij}\hat{a}_a^\dag \hat a_j \hat a_i+ \dots ~ = \hat R_{1h}+\hat R_{2h1p}+\dots ~ 
\end{equation}
The ionized states are then obtained by solving the corresponding EOM equations
\begin{equation}
{[\hat{H}_N,\hat R ]} \ket{\textrm{pCCD}} =  \omega \hat R \ket{\textrm{pCCD}},
\end{equation}
where $ \omega = \Delta E - \Delta E_0$ is the energy difference associated with the ionization process with respect to the pCCD ground state,
while $\hat{H}_N = \hat{H} - \langle\Phi_0\big|\hat{H}|\Phi_0\rangle  $ is the normal product form of the Hamiltonian.
The above equation can be rewritten as
\begin{equation}
{\cal H}^\textrm{pCCD}_N \hat R\ket{\Phi_0} =  \omega \hat R \ket{\Phi_0}
\end{equation}
with ${\cal H}^\textrm{pCCD}_N$ being the similarity transformed Hamiltonian of pCCD in its normal-product form ${\cal H}^\textrm{pCCD}_N$= $e^{- \hat{T}_\textrm{pCCD}}\hat{H}_Ne^{\hat{T}_\textrm{pCCD}}$.
The ionization energies are thus the eigenvalues of a non-Hermitian matrix, which can be iteratively diagonalized to determine the lowest-lying ionized states.

So far, the IP-EOM-pCCD model has been used to study a variety of electronic properties of small organic molecules, including open-shell states and singlet-triplet gaps.
However, a thorough benchmark study for ionization potentials predicted by the IP-EOM-pCCD model has not been presented, yet, and it remains inconclusive whether ionization potentials can be accurately determined by restricting the reference wavefunction ansatz to the electron-pair sector.
In this work, we evaluate the accuracy and reliability of the inexpensive and simple IP-EOM-pCCD model to predict ionization energies in small organic compounds and compare it to more elaborate approaches.

This paper is organized as follows. In section \ref{sec:comp} we describe the methodology and benchmarks used in this study of the electronic properties of small organic molecules. Numerical results and a statistical analysis are presented in section \ref{sec:results}. Finally, we conclude in section \ref{sec:conclusion}.

\begin{figure}[t]
\centering
  \includegraphics[width=100.0mm,scale=0.5]{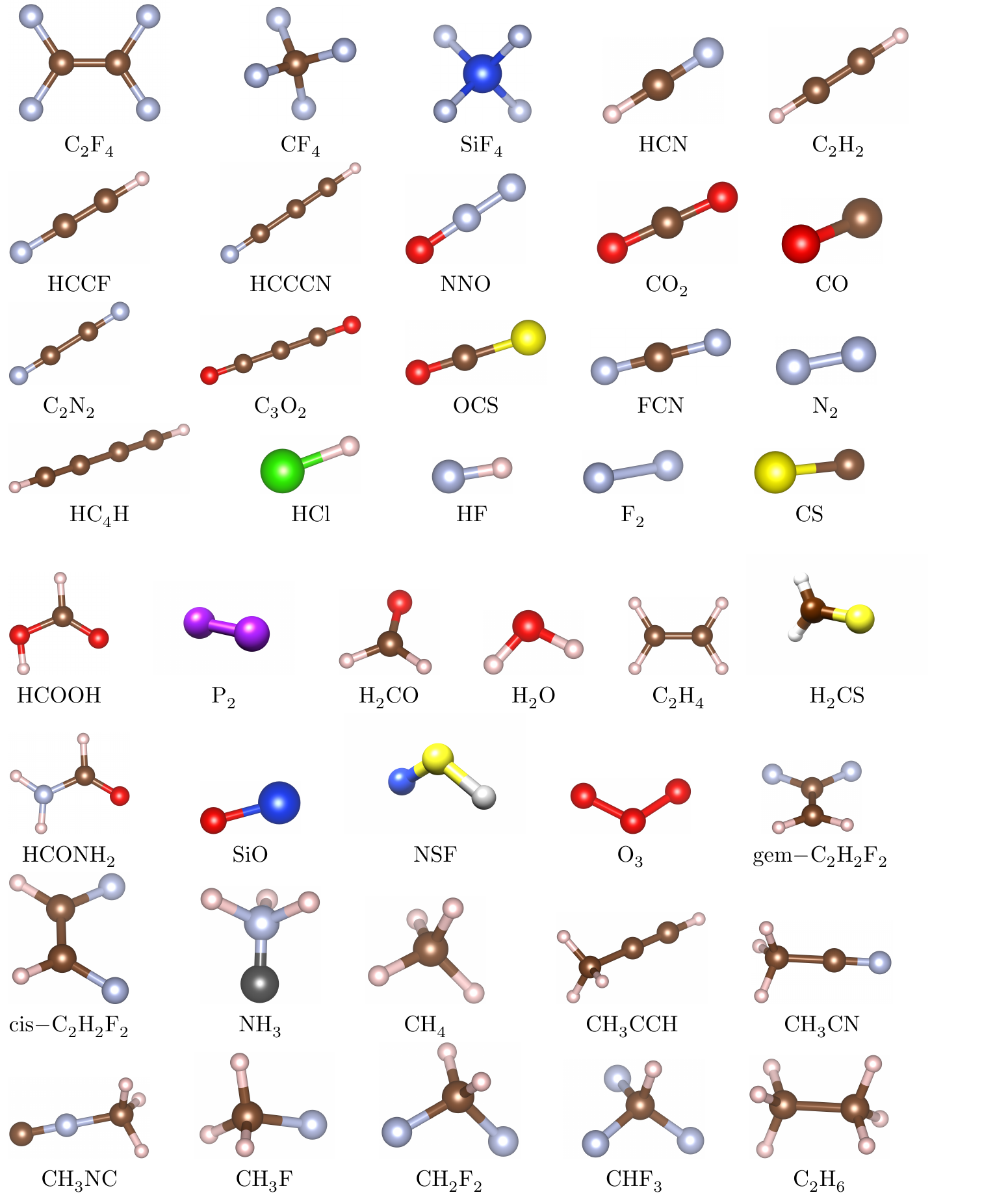}
  \caption{The benchmark set containing 41 molecules relaxed at the CCSD(T)/aug-cc-pVTZ level provided in Ref.~\citenum{ranasinghe2019vertical}.} 
  \label{fgr:molecules}
\end{figure}

\begin{table}[b!]
\caption{Statistical error measures [eV], including mean error (ME), mean absolute error (MAE), and root-mean-square error (RMSE), evaluated from the IP values calculated with the IP-EOM-pCCD method (upper part) and selected quantum chemistry methods (lower part) with respect to IP-EOM-CCSDT~\cite{ranasinghe2019vertical} (top) and experimental~\cite{ranasinghe2019vertical} (bottom) reference data.
ME, MAE, and RMSE are defined in the table footnote.
The IP-EOM-pCCD calculations were performed using three different types of molecular orbitals, including canonical restricted Hartree-Fock (RHF) orbitals, Pipek-Mezey localized (PM) orbitals, and natural pCCD-optimized (pCCD) orbitals, and calculated in the space of the one-hole ({1h}) and two-hole-one-particle ({2h1p}) states.
The corresponding errors in IPs obtained for the CCSD, unitary coupled cluster (UCC), and algebraic-diagrammatic construction (ADC) methods are given in the lower part.
All IP-EOM-CC-type methods are abbreviated as IP, dropping the EOM part.}
\label{tab:errors}
\begin{tabular*}{0.48\textwidth}{@{\extracolsep{\fill}}lccc}
\hline
 & \multicolumn{3}{c}{Errors w.r.t.~IP-EOM-CCSDT} \\\cline{2-4}
Method & ME & MAE &  RMSE\\
\hline
 IP-pCCD(1h), RHF           & \PM2.021  & 2.021 & 2.261\\
 IP-pCCD(1h), PM            & \PM2.049  & 2.055 & 2.320\\
 IP-pCCD(1h), pCCD          & \PM2.486  & 2.486 & 2.695\\
 IP-pCCD(2h1p), RHF         & $-$2.144  & 2.144 & 2.210\\
 IP-pCCD(2h1p), PM          & $-$2.155  & 2.155 & 2.233\\
 IP-pCCD(2h1p), pCCD        & $-$1.535  & 1.535 & 1.633\\
 \hline
IP-CCSD\cite{ranasinghe2019vertical}         & \PM0.186 & 0.241 & 0.341\\
IP-UCC2\cite{dempwolff2022vertical}         & $-$0.488 & 0.579 & 0.728\\
IP-UCC3\cite{dempwolff2022vertical}         & \PM0.260 & 0.306 & 0.377\\
IP-ADC(2)\cite{dempwolff2022vertical}       & $-$0.545 & 0.607 & 0.737\\
IP-ADC(3(3))\cite{dempwolff2022vertical}    & \PM0.269 & 0.351 & 0.442\\
IP-ADC(3(4+))\cite{dempwolff2022vertical}   & \PM0.306 & 0.339 & 0.418\\
IP-ADC(3(DEM))\cite{dempwolff2022vertical}  & \PM0.292 & 0.334 & 0.411\\
\hline
 & \multicolumn{3}{c}{Errors w.r.t.~experiment} \\\cline{2-4}
Method & ME & MAE &  RMSE\\
\hline
IP-pCCD(1h), RHF    & \PM1.986 & 1.986 & 2.233\\
IP-pCCD(1h), PM     & \PM2.013 & 2.022 & 2.290\\
IP-pCCD(1h), pCCD   & \PM2.450 & 2.450 & 2.664\\
IP-pCCD(2h1p), RHF  & $-$2.179 & 2.179 & 2.264\\
IP-pCCD(2h1p), PM   & $-$2.190 & 2.190 & 2.286\\
IP-pCCD(2h1p), pCCD & $-$1.570 & 1.570 & 1.697\\ \hline
IP-CCSD\cite{ranasinghe2019vertical}        & \PM0.150 & 0.252 & 0.351\\
IP-CCSDT\cite{ranasinghe2019vertical}       & $-$0.035 & 0.197 & 0.262\\
IP-UCC2\cite{dempwolff2022vertical}        & $-$0.523 & 0.684 & 0.841\\
IP-UCC3\cite{dempwolff2022vertical}        & \PM0.225 & 0.312 & 0.394\\
IP-ADC(2)\cite{dempwolff2022vertical}      & $-$0.580 & 0.693 & 0.844\\
IP-ADC(3(3))\cite{dempwolff2022vertical}   & \PM0.233 & 0.356 & 0.419\\
IP-ADC(3(4+))\cite{dempwolff2022vertical}  & \PM0.271 & 0.342 & 0.423\\
IP-ADC(3(DEM))\cite{dempwolff2022vertical} & \PM0.256 & 0.339 & 0.419\\
\hline
\end{tabular*}
\begin{tablenotes}
   \item[*] ME = $\sum_i^N \frac{E_i^{\rm method} - E_i^{\rm ref}}{N} $
   \item[*] MAE = $\sum_i^N \frac{|E_i^{\rm method} - E_i^{\rm ref}|}{N} $
   \item[*] RMSE = $\sqrt{\sum_i^N \frac{(E_i^{\rm method} - E_i^{\rm ref})^2}{N}}$
\end{tablenotes}
\end{table}

\section{COMPUTATIONAL DETAILS}\label{sec:comp}
The vertical ionization potentials (IP) were calculated using the IP-EOM-pCCD method as implemented in a developer version of the PyBEST software package.\cite{boguslawski2021pythonic}
The optimization protocol used in pCCD calculations automatically selects the reference determinant according to the natural occupation numbers.
No symmetry constraints were imposed.
In all pCCD-based calculations, we introduced a frozen core as follows: for C, N, O, and F, the 1s orbitals were kept frozen, for Si, P, S, and Cl, we froze the 1s, 2s, and 2p orbitals.
We should note that selected test calculations indicate that freezing the core orbitals only marginally affects the IP values.
The influence of the type of molecular orbitals used during the pCCD optimization procedure on the IP values was investigated using (i) canonical restricted Hartree-Fock orbitals (labeled as RHF), (ii) orbitals localized using the Pipek-Mezey (labeled as PM) scheme,~\cite{pipek1989fast} and (iii) natural pCCD-optimized orbitals (labeled as pCCD) employing a variational orbital-optimization protocol.~\cite{boguslawski2014efficient,boguslawski2014nonvariational,boguslawski2014projected}
The ionization energies were calculated in the space of the one-hole ({1h}) and two-hole-one-particle ({2h1p}) states. 

All calculations utilized the cc-pVTZ basis set of Dunning\cite{dunning1989gaussian} to ensure direct comparison with the previously published vertical IPs calculated with the IP-EOM-CCSDT model,~\cite{ranasinghe2019vertical} different variants of the unitary coupled-cluster (IP-UCC) flavor and the algebraic-diagrammatic construction (IP-ADC) methods.\cite{dempwolff2022vertical}
Since the influence of effects beyond triple excitations is neglected in IP-EOM-CC methods, we limit our discussion in this work to a comparison with IP-EOM-CCSDT reference data.~\cite{ranasinghe2019vertical}

Furthermore, our test set contains 41 molecules shown in Fig.~\ref{fgr:molecules}, whose molecular geometries relaxed at the CCSD(T)/aug-cc-pVTZ level~\cite{kendall1992electron} are available in the supplementary material of Ref.~\citenum{ranasinghe2019vertical}.
In total, we optimized 201 IP states and compared the performance of IP-EOM-pCCD to IP-EOM-CCSDT and experimental reference data.

\section{RESULTS AND DISCUSSION}\label{sec:results}

\begin{figure*}
\centering
\includegraphics[width=0.9\textwidth]{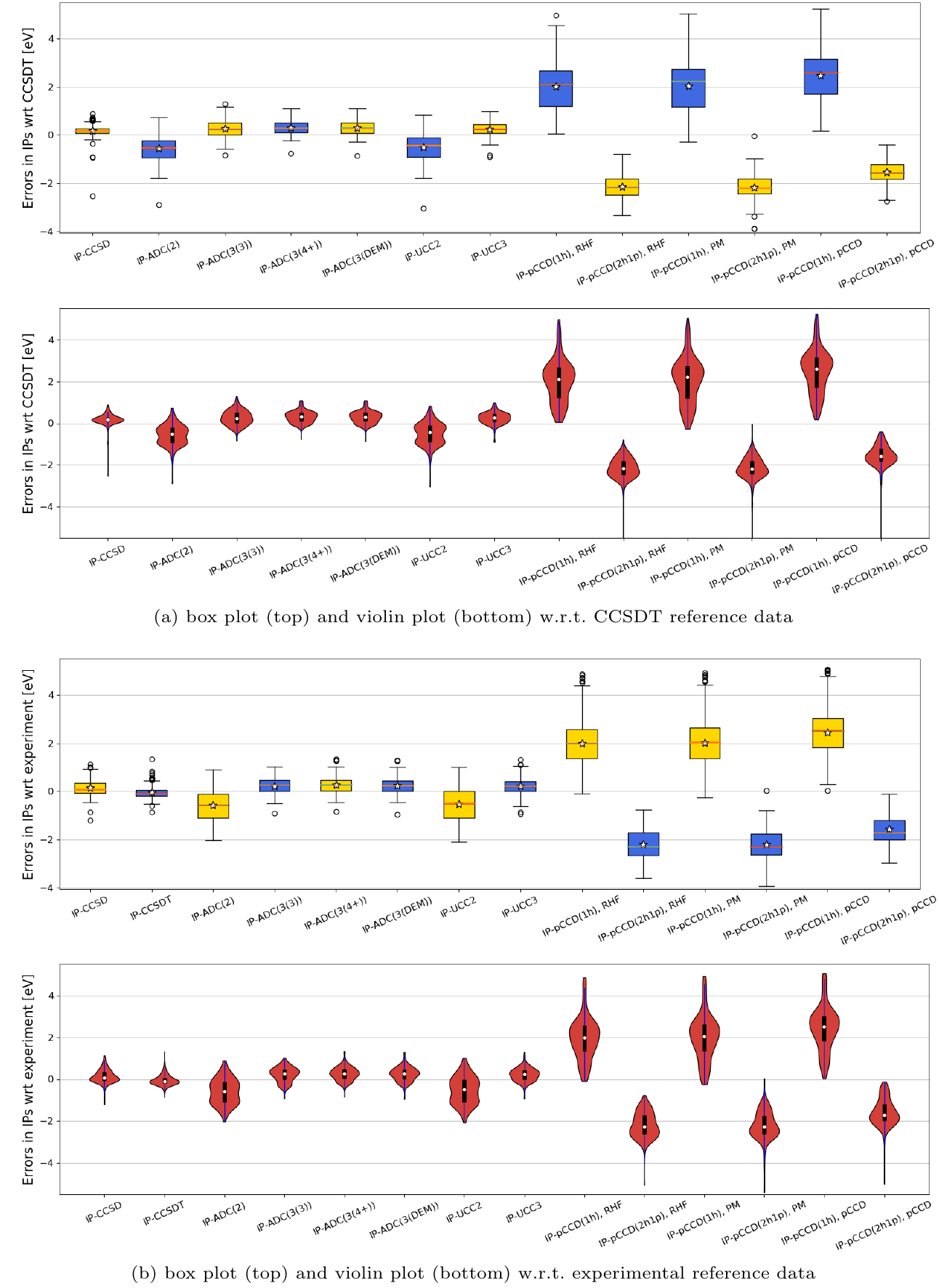}
\caption{Box plot (top) and violin plot (bottom) of errors [eV] obtained with some selected methods (for numerical values see Table 1 and 2). All errors are given with respect to (a) IP-EOM-CCSDT or (b) experimental reference data. Note that one outlier for IP-pCCD(2h1p) is not shown in the plots and is located around $-$6.2 eV. We have dropped the EOM prefix in IP-EOM-CC-type methods for reasons of brevity. The star in each box plot indicates the mean value.}
\label{fig:errorplots}
\end{figure*}

Table~\ref{tab:errors} summarizes all statistical measures in terms of the mean error (ME), mean absolute error (MAE), and root-mean-square error (RMSE) evaluated from the 201 ionized states in 41 molecules calculated by various IP-EOM-pCCD flavors (cf.~Fig.~\ref{fgr:molecules} for the molecular test set and the footnote in Table~\ref{tab:errors} for the definition of the error measures).
The upper part of the Table collects all error values with respect to IP-EOM-CCSDT reference data,~\cite{ranasinghe2019vertical} while the lower part gathers the corresponding errors with respect to experimental results.~\cite{ranasinghe2019vertical}
For a direct comparison, we additionally list the data obtained from IP-EOM-CCSD, two variations of the unitary coupled cluster ansatz (IP-UCC2 and IP-UCC3), and four flavors of the non-Dyson algebraic diagrammatic construction schemes (ADC(2), ADC(3(3), ADC(3(4+)), and ADC(3(DEM)).\cite{dempwolff2022vertical}
Furthermore, Fig.~\ref{fig:errorplots} highlights a graphical representation of the performance of various quantum chemistry methods using box and violin plots of the predicted IPs of all investigated approaches.
Specifically, Fig.~\ref{fig:errorplots}(a) visualizes the locality, spread, skewness, and distribution of the errors in IPs with respect to IP-EOM-CCSDT, while Fig.~\ref{fig:errorplots}(b) illustrates an equivalent analysis with respect to experimental data.
We should note that the pCCD-based calculations were performed with three different types of molecular orbitals (canonical RHF, PM, and orbital-optimized pCCD) including only {1h} and also {2h1p} states in the IP ansatz.
The individual ionization energies obtained by all investigated IP-EOM-pCCD methods and their correlation plots resolved for each molecule, molecular orbital basis, and IP-EOM ansatz are available in the ESI.

In general, IP-EOM-pCCD(1h) (an IP ansatz restricted to a 1h operator) tends to overestimate all IPs compared to the IP-EOM-CCSDT and experimental reference values, regardless of the type of molecular orbitals used (cf.~Table~\ref{tab:errors}).
Among this set of results (that is, restricted to 1h), IP-EOM-pCCD with canonical RHF orbitals features the smallest error measures, reducing them by about 0.4 eV.
However, RHF and PM-localized orbitals yield almost identical error measures, differing only by 0.028 (0.027) eV for ME, 0.034 (0.036) eV for MAE, and 0.059 (0.057) eV for RMSE estimated with respect to IP-EOM-CCSDT (experiment).
Somewhat larger errors are obtained from IP-EOM-pCCD(1h) using the natural pCCD-optimized orbitals.
Nonetheless, the box plots in Fig.\ref{fig:errorplots} demonstrate that the dispersion of 50\% of errors (marked by blue/yellow boxes) and the total range of scores (marked by the whiskers) are similar for all IP-EOM-pCCD(1h) methods.
Additionally, the distribution of all IP-EOM-pCCD(1h) errors (cf.~the violin plots in Fig.~\ref{fig:errorplots}) is concentrated up to 75\%, with their maximum around the mean values.
Thus, all investigated IP-EOM-pCCD(1h) flavors yield similar error distributions, suggesting that all studied molecular orbital basis sets produce qualitatively similar IPs.
Finally, an orbital optimization within pCCD is generally not recommended if the IP ansatz is restricted to 1h operators.

In contrast to IP-EOM-pCCD(1h), the 2h1p analog IP-EOM-pCCD(2h1p) consistently underestimates ionization energies compared to both the IP-EOM-CCSDT and experimental reference data (see also the correlation plots summarized in the ESI).
Furthermore, a change in the molecular orbital basis more strongly affects the performance of pCCD-based methods.
Most importantly, employing natural pCCD-optimized orbitals and extending the IP ansatz to include 2h1p operators significantly improves ionization energies compared to its 1h counterpart IP-EOM-pCCD(1h), reducing the MAE and RMSE errors by 0.951 (0.452) eV and 1.062 (0.593) eV, respectively, compared to the IP-EOM-CCSDT (experimental) reference data.
Furthermore, the errors with respect to IP-EOM-CCSDT (experimental) data change only slightly, after imposing the {2h1p} operator in the IP ansatz and employing canonical RHF or PM-localized molecular orbitals: $\Delta$MAE=0.123 (0.204) eV and $\Delta$RMSE=$-0.051$ (0.053) eV for RHF orbitals, and $\Delta$MAE = 0.100 (0.271) eV and $\Delta$RMSE = $-0.087$ (0.400) eV for PM-localized orbitals.
Finally, introducing the {2h1p} operator decreases the dispersion of the data and significantly shifts the distribution of errors towards the mean values/median when the molecular orbitals are optimized within pCCD (cf.~Fig.~\ref{fig:errorplots}).

Our statistical analysis suggests that IP-EOM-pCCD(2h1p) employing natural pCCD-optimized orbitals is the best method among all investigated IP-EOM-pCCD approaches.
However, its errors are unacceptably worse than those obtained by the IP-UCC2 and IP-ADC(2) methods (varying by a factor of 2 to 3 with respect to the choice of reference values and the type of the error measure), which are the least elaborated approximations of the UCC and ADC formalism.
In contrast to IP-UCC2 and IP-ADC(2), IP-EOM-pCCD(2h1p) features an error distribution that is more centered around the mean value of the distribution, while the former two predict a dumbbell-shaped distribution of errors, which are more scattered.
Dissecting the ionization energies for the individual molecules, the majority of compounds, besides some pathological cases as O$_3$, P$_2$ or C$_2$N$_2$, show a linear trend for the ionization potentials, which are shifted to lower ionization energy values (see the correlation plots in the ESI).
These noticeable deviations can be attributed to the missing dynamical electron correlation effects that cannot be described by pCCD and are associated with broken-pair states.
Nevertheless, IP-EOM-pCCD(2h1p) exploiting natural pCCD-optimized orbitals exhibits a similar trend with respect to the error dispersion and a much better error distribution than IP-UCC2 and IP-ADC(2).

\section{CONCLUSIONS}\label{sec:conclusion}
In this paper, we scrutinized the performance of the orbital-optimized pCCD ansatz in predicting vertical valence ionization potentials using the IP-EOM formalism recently introduced.~\cite{boguslawski2021open}
Specifically, we benchmarked IP-EOM-pCCD for three different sets of molecular orbitals (canonical restricted Hartree--Fock, Pipek--Mezey localized, and natural pCCD-optimized molecular orbitals) and two sets of particle-hole operators (1h and 2h1p) against a test set of 41 organic molecules targeting 201 ionized states.
The IP-EOM-pCCD ionization energies were compared to IP-EOM-CCSDT and experimental reference data.
In general, the errors of IP-EOM-pCCD(2h1p) feature a similar spread and skewness (derived from box plots) as the more sophisticated IP-ADC(2) and IP-UCC2 methods, while the error distributions (derived from violin plots) is centered around the median, in contrast to IP-ADC(2) and IP-UCC2, which show a dumbbell-shaped distribution of errors.
Nonetheless, the ME and RSME of IP-EOM-pCCD deviate considerably from IP-EOM-CCSDT and experimental reference data, spreading from 1.5 (2h1p) to 2.6 eV (1p).
The best performance was obtained for the natural pCCD-optimized molecular orbital basis and including the 2h1p operator in the IP ansatz, lowering the ME and RMSE to 1.5 eV and 1.6 eV respectively.
Thus, IP-EOM-pCCD(2h1p) results in ionization potentials that are reasonably spread but considerably shifted due to the missing dynamical correlation effects that are not included in the pCCD reference function and cannot be accounted for in the IP ansatz.

In summary, IP-EOM-pCCD represents an extremely simple model that is able to target open-shell states starting from the closed-shell pCCD reference function and a panacea for a computationally cheap and robust alternative to more elaborate quantum chemistry methods.
Regardless of its drawbacks, its performance can be further improved by including dynamical correlation on top of the pCCD reference function, for instance using the frozen pCC formalism~\cite{henderson2014seniority} or its linearized variant,~\cite{boguslawski2015linearized} which is currently under investigation in our laboratory.

\section*{Acknowledgements}
The research leading to these results has received funding from the Norway Grants 2014--2021 via the National Centre for Research and Development.
The research leading to these results has received funding from the Norway Grants 2014--2021 via the National Centre for Research and Development.
M.G.~acknowledges financial support from a Ulam NAWA -- Seal of Excellence research grant (no.~BPN/SEL/2021/1/00005). 


\renewcommand\refname{References}

\bibliography{rsc} 
\bibliographystyle{rsc} 

\end{document}